\documentclass[11pt]{article}
\usepackage{amsmath, amsthm, amsfonts}
\usepackage[margin=1in,footskip=0.25in]{geometry}
\usepackage{makecell}
\theoremstyle{theorem}

\newtheoremstyle{defi}%name
  {10pt}          % space above
  {10pt}  % space below
  {\rm}  % bofy font
  {\parindent}     % ident - empty=no indent,  \parindent= paragraph indent
  {\bf}  % thm head font
  {. }    % punctuation after thm head
  { }    % space after thm head: `` ``=normal \newline=linebreak
  {}     % thm head specification
\theoremstyle{defi}

\begin{document}

\date{}

\title{\bf Imperfect fluid cosmological model in modified gravity}
\author{G. C. Samanta$^{1}$, R. Myrzakulov$^{2}$\\
 $^{1}$Department of Mathematics\\
Birla Institute of Technology and Science (BITS) Pilani,\\
K K Birla Goa Campus,\\
Goa-403726, India,\\ gauranga81@gmail.com\\
$^{2}$Eurasian International Center for Theoretical Physics \\ and Department of General Theoretical
Physics, \\Eurasian National University, Astana 010008, Kazakhstan\\
rmyrzakulov@gmail.com}
%$^{2.}$ Department of Mathematics\\
%Birla Institute of Technology and Science-Pilani,\\
%Hyderabad Campus,\\
%Hyderabad-500078, INDIA,\\ bivudutta@yahoo.com}

\maketitle

\begin{abstract} In this article, we considered the bulk viscous fluid in the formalism of modified gravity in which the general form of a gravitational action is $f(R, T)$ function, where $R$ is the curvature scalar and $T$ is the trace of the energy momentum tensor within the frame of flat FRW space time. The cosmological model dominated by bulk viscous matter with total bulk viscous coefficient expressed as a linear combination of the velocity and acceleration of the expansion of the universe in such a way that $\xi=\xi_0+\xi_1\frac{\dot{a}}{a}+\xi_2\frac{\ddot{a}}{\dot{a}}$, where $\xi_0$, $\xi_1$ and $\xi_2$ are constants. We take $p=(\gamma-1)\rho$, where $0\le\gamma\le2$ as an equation of state for perfect fluid. The exact solutions to the corresponding field equations are obtained by assuming a particular model of the form of $f(R, T)=R+2f(T)$, where $f(T)=\lambda T$, $\lambda$ is constant. We studied the four possible scenarios of the universe for different values of $\gamma$, such as $\gamma=0$, $\gamma=\frac{2}{3}$, $\gamma=1$ and $\gamma=\frac{4}{3}$ with the possible positive and negative ranges of $\lambda$ to observe the accelerated expansion history of the universe. Finally, a big-rip singularity is observed.
\end{abstract}

%\textbf{AMS Classification number}: 83F05\\


\section{Introduction}
  From the type Ia supernova observations, it is clear that the present universe is dominated by dark energy which provides the
  dynamical mechanism of the accelerated expansion of the universe (\cite{riess}, \cite{Garnavich}, \cite{perlmutter}). Further it was confirmed by the observations from Cosmic Microwave
  Background Radiations (CMBR) \cite{Bennet 2003}, Large Scale Structure (LSS) \cite{tegmark2004}, the Sloan Digital Sky Survey (SDSS) \cite{Seljan2005}, the Wilkinson Microwave Anisotropy Probe (WMAP) \cite{komatsu2011} etc. The strength of this acceleration is a remarkable question in recent year. There are many models have been introduced to explain
  this current acceleration of the universe. Generally there are two approaches to describe the current acceleration of the universe: one is to propose
  to modify the energy momentum tensor $T_{\mu\nu}$ in the Einstein's field equations. The second approach is to modify the geometry of the space time
  in the Einstein's equations.
  \par
  The simplest candidate for dark energy is the cosmological constant $(\wedge)$, which is so called because its energy density is constant with respect
to time and space. However, it suffers from the coincidence problem and the fine tuning problem \cite{copeland2006}. So as a result, dynamical dark energy models such
as quintessence (\cite{Fujii1982}, \cite{carroll1998}), k-essence (\cite{chiba2000}, \cite{armendariz2000}) and perfect fluid models (like Chaplygin gas model) (\cite{kamenshchik2001}, \cite{bento2002}) were considered.
\par
Recently, the modified gravity has become one of the most popular candidates to understand the idea of dark energy. In modified gravity, the origin of dark
energy is identified as a modification of gravity. In literature, a number of modified theories have been discussed to explain early and late time expansion of the universe. In modified theories one modifies the laws of gravity so that the late time accelerated expansion of the universe is realized
without recourse to an explicit dark energy matter component. One of the simplest modified gravity model is the so-called $f(R)$ gravity in which the
4-dimensional action is given by some general function $f(R)$ of the Ricci scalar $R$
\begin{equation}\label{1}
S=\frac{1}{2k^2}\int d^4x\sqrt{-g}f(R)+S_m(g_{\mu\nu}, \psi_{m})
\end{equation}
where $k^2=8\pi G$ and $S_m$ is a matter action with matter field $\psi_m$. The matter field in $S_m$ obey standard conservation equations. The $f(R)$
gravity was first introduced by \cite{buchdahl1970}. Subsequently, several authors (\cite{starobinsky1980}, \cite{nojiri2003}, \cite{sotiriou2010}, \cite{capozziello2001}, \cite{chifton2012}, \cite{Snojiri2003}, \cite{Nojiri2004}, \cite{nojiri2011}, \cite{carroll2004}, \cite{chiba2007}, \cite{bamba2008}, \cite{bamba2012}, \cite{starobinsky2007}) investigated $f(R)$ theory of gravity from different aspects. However, there are
many modified theories of gravity have been developed like $f(T)$ gravity, Gauss-Bonnet theory, Lovelock gravity, Horva-Lifshitz gravity, scalar-tensor
theories of gravity, braneworld models and so on (\cite{ferraro2007}, \cite{myrzakulov2011}, \cite{nojiri2005}, \cite{padmanabhan2013}, \cite{horava2009}, \cite{Amendola1999}, \cite{Dvali2000}, \cite{janil2012}).
\par
Recently, \cite{harko2011} developed an another modification of general relativity, so called the $f(R, T)$ gravity, where the gravitational Lagrangian is given by an
arbitrary function of the curvature scalar $R$ and $T$ is the torsion scalar. The $f(R, T)$ gravity model depends on a source term, representing the
variation of the matter stress-energy tensor with respect to the metric. The general expression for this source term is obtained as a function of the matter Lagrangian $L_m$. Therefore, the different choice of $L_m$ would generate a specific set of field equations. This modified theory possessing
some interesting results which are relevant in theoretical cosmology and astrophysics. In most of the cosmological models, the matter part of the
universe has been considered as a perfect fluid. In the context of inflation, many authors investigated that the bulk viscous fluids are capable of
providing acceleration of the universe (\cite{padmanabhan1987}, \cite{wega1986}, \cite{cheng1991}). The bulk viscous fluid is the unique viscous effect capable to modify the background dynamics in a homogeneous
and isotropic universe. The matter behave like a viscous fluid in the early stage of the universe in neutrino decoupling phase. It has been known
that, perfect fluid with bulk viscosity can produce an acceleration without the help of a cosmological constant or some scalar field. This idea was
extended to explain the late time acceleration of the universe (\cite{fabris2006}, \cite{li2009}, \cite{hipolito2010}, \cite{avelino2009}, \cite{avelino2010}). Inhomogeneous equation of state of the universe, such as phantom era, future singularity and crossing the phantom barrier discussed by \cite{nojiri47}. The effects of viscosity terms depending on the Hubble parameter and its derivatives in the dark energy equation of state discussed by \cite{Capozziello48}. Brevik et al.\cite{Brevik49} proved, in particular, that a viscous fluid (or, equivalently, one with an inhomogeneous (imperfect) equation of state) is perfectly able to produce a Little Rip cosmology as a purely viscosity effect. The bulk viscosity is the most favorable phenomenon, compatible with the symmetry
requirements of the homogeneous and isotropic universe. The dark energy phenomenon as an effect of the bulk viscosity has been investigated in \cite{zimdahl2001}.
The mechanism for the formation of bulk viscosity by the decay of a dark matter particle is discussed in \cite{wilson2007}.
\par
The motivation of the present work is to discuss the present acceleration of the universe by the help of imperfect fluid within the framework of
$f(R, T)$ gravity. In this paper, authors studied homogeneous and isotropic universe with bulk viscosity in $f(R, T)$ gravity and discussed the
effects of bulk viscosity in explaining the early and late time acceleration of the universe. The authors analysed the cosmic evolution of the bulk
viscous matter dominated universe with the bulk viscous coefficient $\xi$ depending on both the velocity and acceleration of the
expanding universe as, $\xi=\xi_0+\xi_1 \frac{\dot{a}}{a}+\xi_2 \frac{\ddot{a}}{a}$, where $a$ is the scale factor of the universe and
$\xi_0, \xi_1 ~\mbox{\&} ~\xi_2$ are constants. The exact solutions of the field equations are obtained by assuming a simplest form of $
f(R, T)=R+2f(T)$, where $f(T)=\lambda T$.

\section{Brief review of f(R, T) gravity and field equations}
The $f(R, T)$ theory is a modification of Einstein-general theory of relativity, in which the Einstein-Hilbert Lagrangian, i. e. $R$ is replaced by
an arbitrary function of the curvature scalar $R$ and the trace $T$ of energy momentum tensor. The following modification of Einstein theory is proposed by (\cite{harko2011}). \\
The action for the modified gravity takes the following form
\begin{equation}\label{2}
  S=\frac{1}{16 \pi}\int\left(f(R, T)+16\pi L_m\right)\sqrt{-g}d^4x
\end{equation}
where $g$ is the determinate of the metric tensor $g_{\mu\nu}$ and $L_m$ is the matter Lagrangian density. \\
The energy momentum tensor $T_{\mu\nu}$, defined from the matter Lagrangian density $L_m$ is given by
\begin{equation}\label{3}
T_{\mu\nu}=\frac{-2}{\sqrt{-g}}\frac{\delta(\sqrt{-g}L_m)}{\delta g^{\mu\nu}}
\end{equation}
 and its trace $T$ is defined by $T=g^{\mu\nu}T_{\mu\nu}$. \\
 Assume, the Lagrangian density $L_m$ of matter depends only in the metric tensor $g_{\mu\nu}$, not on its derivatives, can obtain
 \begin{equation}\label{4}
   T_{\mu\nu}=g_{\mu\nu}L_m-2\frac{\partial L_m}{\partial g^{\mu\nu}}.
 \end{equation}
The variation of the action \eqref{2} with respect to $g^{\mu\nu}$ gives
\begin{eqnarray} \label{5}
  \delta S &=& \frac{1}{16\pi}\int\bigg[f_R(R, T)\delta R+f_T(R, T)\frac{\delta T}{\delta g^{\mu\nu}}\delta g^{\mu\nu}
  -\frac{1}{2}g_{\mu\nu}f(R, T)\delta g^{\mu\nu} \nonumber \\
   &+& 16\pi \frac{1}{\sqrt{-g}}\frac{\delta (\sqrt{-g}L_m)}{\delta g^{\mu\nu}}\bigg]\sqrt{-g}d^4x
\end{eqnarray}
where $f_R(R, T)=\frac{\partial f(R, T)}{\partial R}$ and $f_T(R, T)= \frac{\partial f(R, T)}{\partial T}$, respectively.
The variation of the Ricci scalar can be obtained as
\begin{eqnarray} \label{6}
  \delta R &=& \delta(g^{\mu\nu}R_{\mu\nu}) \nonumber \\
  &=& R_{\mu\nu}\delta g^{\mu\nu}+g^{\mu\nu}(\bigtriangledown_{\lambda}\delta \Gamma_{\mu\nu}^{\lambda}-\bigtriangledown_{\nu}\delta T_{\mu\lambda}^{\lambda}),
\end{eqnarray}
where $\bigtriangledown_{\lambda}$ is the covariant derivative with respect to the symmetric connection is associated to the metric $g$. The
variation of the Christoffel symbols yields
\begin{equation}\label{7}
  \delta \Gamma_{\mu\nu}^{\lambda}=\frac{1}{2}g^{\lambda\alpha}(\bigtriangledown_{\mu}\delta g_{\nu\alpha}+\bigtriangledown_{\nu}
  \delta g_{\alpha\mu}-\bigtriangledown_{\alpha}\delta g_{\mu\nu}),
\end{equation}
and the variation of the Ricci scalar perform
\begin{equation}\label{8}
  \delta R=R_{\mu\nu}\delta g^{\mu\nu}+g_{\mu\nu}\Box \delta g^{\mu\nu}-\bigtriangledown_{\mu}\bigtriangledown_{\nu}\delta g^{\mu\nu}.
\end{equation}
Thus \eqref{5} reduces to
\begin{eqnarray} \label{9}
  \delta S &=& \frac{1}{16\pi}\int\bigg[f_R(R, T)R_{\mu\nu}\delta g^{\mu\nu}+f_R(R, T)g_{\mu\nu}\Box\delta g^{\mu\nu}\nonumber \\
   &-& f_R(R, T)\bigtriangledown_{\mu}\bigtriangledown_{\nu}\delta g^{\mu\nu}+f_T(R, T)
   \frac{\delta(g^{\alpha\beta }T_{\alpha\beta})}{\delta g^{\mu\nu}}\delta g^{\mu\nu} \nonumber \\
  &-& \frac{1}{2} g_{\mu\nu}f(R, T) \delta g^{\mu\nu}+16\pi \frac{1}{\sqrt{-g}}\frac{\delta(\sqrt{-g}L_m)}{\delta g^{\mu\nu}}\bigg]
  \sqrt{-g}d^4x.
\end{eqnarray}
The variation of $T$ with respect to the metric tensor is obtained as
\begin{equation}\label{10}
  \frac{\delta(g^{\alpha\beta}T_{\alpha\beta})}{\delta g^{\mu\nu}}=T_{\mu\nu}+\Theta_{\mu\nu}
\end{equation}
where
\begin{equation}\label{11}
  \Theta \equiv g^{\alpha\beta}\frac{\delta T_{\alpha\beta}}{\delta g^{\mu\nu}}.
\end{equation}
After integrating the second and third terms in equation \eqref{9}, we obtained the field equations of the $f(R, T)$ gravity model as
\begin{eqnarray} \label{12}
  f_R(R, T)R_{\mu\nu}&-&\frac{1}{2}f(R, T)g_{\mu\nu}+(g_{\mu\nu}\Box-\bigtriangledown_{\mu}\bigtriangledown_{\nu})f_R(R, T)\nonumber  \\
  &=&8\pi T_{\mu\nu}-f_T(R, T)T_{\mu\nu}-f_T(R, T)\Theta_{\mu\nu}.
\end{eqnarray}
where $T_{\mu\nu}$ is the standard matter energy momentum tensor derived from equation \eqref{4}.
\begin{eqnarray} \label{13}
  \frac{\delta T_{\alpha\beta}}{\delta g^{\mu\nu}} &=& \frac{\delta g_{\alpha\beta}}{\delta g^{\mu\nu}}L_m+g_{\alpha\beta}\frac{\partial L_m}{
  \partial g^{\mu\nu}}-2\frac{\partial^2L_m}{\partial g^{\mu\nu}\partial g^{\alpha\beta}} \nonumber \\
  &=& \frac{\delta g_{\alpha\beta}}{\delta g^{\mu\nu}}L_m+ \frac{1}{2}g_{\alpha\beta}g_{\mu\nu}L_m-\frac{1}{2}
  g_{\alpha\beta}T_{\mu\nu}-2\frac{\partial^2L_m}{\partial g^{\mu\nu}\partial g^{\alpha\beta}}
\end{eqnarray}
From the condition $g_{\alpha\sigma}g^{\sigma\beta}=\delta_{\alpha}^{\beta}$, we have
\begin{equation}\label{14}
  \frac{\delta g_{\alpha\beta}}{\delta g^{\mu\nu}}=-g_{\alpha\sigma}g_{\beta\gamma}\delta_{\mu\nu}^{\sigma\gamma}
\end{equation}
where $\delta_{\mu\nu}^{\sigma\gamma}=\frac{\delta g^{\sigma\gamma}}{\delta g^{\mu\nu}}$ is the generalized kronecker symbol. Therefore, $\Theta_{\mu\nu}$
is defined as
\begin{equation}\label{15}
  \Theta_{\mu\nu}=-2T_{\mu\nu}+g_{\mu\nu}L_m-2g^{\alpha\beta}\frac{\partial^2L_m}{\partial g^{\mu\nu}\partial g^{\alpha\beta}}
\end{equation}
The contraction of equation \eqref{12} yields
\begin{equation}\label{16}
  f_R(R, T)R+3\Box f_R(R, T)-2f(R, T)=8\pi T-f_T(R, T)T-f_T(R, T)\Theta
\end{equation}
where $\Theta=g^{\mu\nu}\Theta_{\mu\nu}$ and $\Box \equiv \bigtriangledown_{\mu} \bigtriangledown^{\mu}$ is the d'Alembert operator. From the
equations \eqref{12} and \eqref{16} we obtain
\begin{eqnarray} \label{17}
  f_R(R, T)(R_{\mu\nu}-\frac{1}{3}Rg_{\mu\nu})+\frac{1}{6}f(R, T)g_{\mu\nu}&=&
  8\pi (T_{\mu\nu}-\frac{1}{3}Tg_{\mu\nu})-f_T(R, T)(T_{\mu\nu}-\frac{1}{3}Tg_{\mu\nu})\nonumber \\
  &-& f_T(R, T)(\Theta_{\mu\nu}-\frac{1}{3}\Theta g_{\mu\nu})+\bigtriangledown_{\mu}\bigtriangledown_{\nu}f_R(R, T).
\end{eqnarray}
If we assume the matter of the universe as a perfect fluid, then the stress energy momentum tensor of the matter Lagrangian is given by
\begin{equation}\label{18}
  T_{\mu\nu}=(p+\rho)u_{\mu}u_{\nu}-pg_{\mu\nu},
\end{equation}
and the matter Lagrangian can be taken as $L_m=-p$. The four velocity vector in co-moving co-ordinates system is defined as
$u^{\mu}=(1, 0, 0, 0)$ which satisfies the conditions $u_{\mu}u^{\mu}=1$ and $u^{\mu} \bigtriangledown_{\nu}u_{\mu}=0$.
Here $p$ and $\rho$ are the pressure and energy density of the perfect fluid respectively. With the use of equation \eqref{15}, we obtain for the
variation of the stress-energy of a perfect fluid as
\begin{equation}\label{19}
  \Theta_{\mu\nu}=-2T_{\mu\nu}-pg_{\mu\nu}
\end{equation}
It is important to note that the field equations in $f(R, T)$ gravity also depend on the physical nature of the matter field through the tensor $\Theta_{\mu\nu}$. Therefore, the $f(R, T)$ theory depending on the nature of the matter source. Here, we can obtain several theoretical models for different choice of $f(R, T)$. Harko et al. considered three different explicit form of $f(R, T)$ as
\begin{equation}\label{20}
  f(R, T)=\begin{cases}
            R+2f(T),  \\
             f_1(R)+f_2(T), \\
            f_1(R)+f_2(R)f_3(T).
                      \end{cases}
\end{equation}
Subsequently several authors (\cite{mahanta2014}, \cite{adhav2012}, \cite{ahmed2014}, \cite{houndjo2012}, \cite{mishra2014}, \cite{Reddy2012}, \cite{samanta2013}, \cite{samanta2013a}, \cite{samanta2013b}, \cite{myrzakulov2012}, \cite{singh2014}) studied some cosmological models in $f(R, T)$ modified gravity for different choice of $f(R, T)$ form various aspects.
\par
In this paper, we consider the following form of $f(R, T)$
\begin{equation}\label{21}
  f(R, T)=R+2f(T),
\end{equation}
i. e. the action is given by the same Einstein Hilbert one plus a function of $T$. The term $2f(T)$ in the gravitational action modifies the
gravitational interaction between matter and curvature scalar $R$. Using equation \eqref{21}, one can re-write the gravitational field
equations defined in \eqref{12} as
\begin{equation}\label{22}
  R_{\mu\nu}-\frac{1}{2}R g_{\mu\nu}=8\pi T_{\mu\nu}-2f^{'}(T)(T_{\mu\nu}+\Theta_{\mu\nu})+f(T)g_{\mu\nu}
\end{equation}
which is considered as the field equation of $f(R, T)$ gravity for the above particular form of $f(R, T)$. Here the prime stands for
derivative of $f(T)$ with respect to $T$.
\par
In this paper, we consider the source of gravitation is a combination of perfect fluid and bulk viscous
fluid. Therefore, the energy momentum tensor takes the form
\begin{equation}\label{23}
  T_{\mu\nu}=(\rho+\bar{p})u_{\mu}u_{\nu}-\bar{p}g_{\mu\nu}
\end{equation}
and
\begin{equation}\label{24}
  \bar{p}=p-3\xi H
\end{equation}
where $\rho$ is the energy density, $\xi$ is the coefficient of bulk viscosity, $\bar{p}$ is effective pressure and $p$ is the proper pressure.
Here $H=\frac{\dot{a}}{a}$ is Hubble parameter, where an over dot stands for derivative with respect to cosmic time t. Hence, the Lagrangian density
may be chosen as $L_m=-\bar{p}$ and the tensor $\Theta_{\mu\nu}$ in \eqref{19} reduces to
\begin{equation}\label{25}
  \Theta_{\mu\nu}=-2T_{\mu\nu}-\bar{p}g_{\mu\nu}
\end{equation}
Now using the equations \eqref{23} and \eqref{25}, the field equation \eqref{22} for bulk viscous fluid is given by
\begin{equation}\label{26}
  R_{\mu\nu}-\frac{1}{2}Rg_{\mu\nu}=8\pi T_{\mu\nu}+2f^{'}(T)T_{\mu\nu}+(2\bar{p}f^{'}(T)+f(T))g_{\mu\nu}
\end{equation}
The field equations \eqref{26} with the particular choice of $f(T)=\lambda T$, where $\lambda$
is constant. By assuming that the metric of the universe is given by the flat FRW metric
\begin{equation}\label{27}
 ds^2=dt^2-a^2[dr^2+r^2(d\theta^2+\sin^2\theta d\phi^2)]
\end{equation}
The gravitational field equations are given by
\begin{equation}\label{28}
  3 \left(\frac{\dot{a}}{a}\right)^2=8\rho+2\lambda(\rho+\bar{p})+\lambda T
\end{equation}
\begin{equation}\label{29}
  2\frac{\ddot{a}}{a}+\left(\frac{\dot{a}}{a}\right)^2=-8\pi\bar{p}+\lambda T
\end{equation}
where $T=\rho-3\bar{p}$.
The equation of continuity is given by
\begin{equation}\label{30}
  \dot{\rho}+3\frac{\dot{a}}{a}(\rho+\bar{p})=0
\end{equation}
In the following section we chose equation of state and bulk viscosity coefficient and try to solve for $H$.

\section{Exact solution of the field equations}
The field equations \eqref{28} to \eqref{30} (by substituting $H=\frac{\dot{a}}{a}$) becomes,
\begin{equation}\label{31}
  3H^2=8\pi\rho+2\lambda(\rho+\bar{p})+\lambda T
\end{equation}
\begin{equation}\label{32}
  2\dot{H}+3H^2=-8\pi\bar{p}+\lambda T
\end{equation}
and
\begin{equation}\label{33}
  \dot{\rho}+3H(\rho+\bar{p})=0.
\end{equation}
Subtract equation \eqref{31} from the equation \eqref{32}, yields
\begin{equation}\label{34}
  2\dot{H}+(8\pi+2\lambda)(p+\rho)-3(8\pi+2\lambda)\xi H=0
\end{equation}
We can choose the equation of state in the following form
\begin{equation}\label{35}
  p=(\gamma-1)\rho
\end{equation}
where $\gamma$ is a constant known as the EoS parameter lying in the range $0\le\gamma\le2$.
We assume the general form of bulk viscous coefficient \cite{ren2006}
\begin{eqnarray} \label{36}
  \xi &=& \xi_0+\xi_1\frac{\dot{a}}{a}+\xi_2\frac{\ddot{a}}{\dot{a}} \nonumber \\
   &=& \xi_0+\xi_1H+\xi_2\left(\frac{\dot{H}}{H}+H\right)
\end{eqnarray}
where $\xi_0,~ \xi_1,~\mbox{and}~ \xi_2$ are constants.
\par
In this paper we consider $\xi_0,~ \xi_1,~\mbox{and}~ \xi_2$ all are non zero, so that the total bulk viscous parameter
$\xi=\xi_0+\xi_1\frac{\dot{a}}{a}+\xi_2\frac{\ddot{a}}{\dot{a}}$, depending on both the velocity and acceleration of the expansion of the
universe. Therefore, the linear combination is more general rather than individual.
Using the equations \eqref{24}, \eqref{35} and \eqref{36} into the equation \eqref{31}, we get
\begin{equation}\label{37}
  \rho=\frac{3H\bigg[(1-\lambda(\xi_1+\xi_2))H-\lambda\xi_2\frac{\dot{H}}{H}-\lambda\xi_0\bigg]}{
  8\pi+4\lambda-\lambda\gamma}
\end{equation}
Using the equations \eqref{35} and \eqref{37} into the equation \eqref{34}, we have
\begin{equation*}
  \bigg[2-3(8\pi+2\lambda)\left(\frac{\gamma\lambda}{(8\pi+4\lambda-\lambda\gamma)}+1\right)\xi_2\bigg]\dot{H}-
\bigg[\frac{3\lambda\xi_0\gamma(8\pi+2\lambda)}{8\pi+4\lambda-\lambda\gamma}+3\xi_0(8\pi+2\lambda)\bigg]H
\end{equation*}
\begin{equation}\label{38}
  +\bigg[\frac{(8\pi+2\lambda)3\gamma}{(8\pi+4\lambda-\lambda\gamma)}(1-\lambda(\xi_1+\xi_2))-
  3(8\pi+2\lambda)(\xi_1+\xi_2)\bigg]H^2=0
\end{equation}
This implies that
\begin{eqnarray}\label{39}
  \dot{H} &=& \frac{\bigg[\frac{3\lambda\xi_0\gamma(8\pi+2\lambda)}{(8\pi+4\lambda-\lambda\gamma)}+3\xi_0(8\pi+2\lambda)\bigg]}{
  \bigg[2-3(8\pi+2\lambda)\left(\frac{\gamma\lambda}{(8\pi+4\lambda-\lambda\gamma)}+1\right)\xi_2\bigg]}H \nonumber \\
  &-&\frac{\bigg[\frac{(8\pi+2\lambda)3\gamma}{8\pi+4\lambda-\lambda\gamma}(1-\lambda(\xi_1+\xi_2))-3(8\pi+2\lambda)(\xi_1+\xi_2)\bigg]}{
  \bigg[2-3(8\pi+2\lambda)\left(\frac{\gamma\lambda}{(8\pi+4\lambda-\lambda\gamma)}+1\right)\xi_2\bigg]}H^2
\end{eqnarray}
Equation \eqref{39} is the form of Bernouli differential equation, solving \eqref{39}, we get
\begin{equation}\label{40}
  H=\frac{k_1e^{k_1t}}{k_2e^{k_1t}+k_3}
\end{equation}
where $k_1=\frac{\bigg[\frac{3\lambda\xi_0\gamma(8\pi+2\lambda)}{(8\pi+4\lambda-\lambda\gamma)}+3\xi_0(8\pi+2\lambda)\bigg]}{
  \bigg[2-3(8\pi+2\lambda)\left(\frac{\gamma\lambda}{(8\pi+4\lambda-\lambda\gamma)}+1\right)\xi_2\bigg]}$,
$k_2=\frac{\bigg[\frac{(8\pi+2\lambda)3\gamma}{8\pi+4\lambda-\lambda\gamma}(1-\lambda(\xi_1+\xi_2))-3(8\pi+2\lambda)(\xi_1+\xi_2)\bigg]}{
  \bigg[2-3(8\pi+2\lambda)\left(\frac{\gamma\lambda}{(8\pi+4\lambda-\lambda\gamma)}+1\right)\xi_2\bigg]}$ and $k_3=k_1c_1$, $c_1$ is a constant of integration.
Using $H=\frac{\dot{a}}{a}$, the scale factor $'a'$ is given by
\begin{equation}\label{41}
  a=k_4(k_2e^{k_1t}+k_3)^{\frac{1}{k_2}}
\end{equation}
where $k_4$ is a constant of integration.
The energy density can be calculated as
\begin{equation}\label{42}
  \rho=\frac{3k_1e^{k_1t}}{k_2e^{k_1t}+k_3}\bigg[(1-\lambda(\xi_1+\xi_2))\frac{k_1e^{k_1t}}{k_2e^{k_1t}+k_3}-\lambda\xi_2
  \frac{k_1^2k_3}{(k_2e^{k_1t}+k_3)^2}-\lambda\xi_0\bigg]\frac{1}{(8\pi+4\lambda-\lambda\gamma)}
\end{equation}
The deceleration parameter is given by
\begin{equation}\label{43}
  q=-1-\frac{k_3}{e^{k_1t}},
\end{equation}
which is depends on cosmic time 't'. It seems that, the bulk viscous fluid also, produces time dependent deceleration parameter ($q$) which describe the transition phases of the universe along with deceleration or acceleration of the universe. In table 1, 2, 3 and 4, we presented the variation of deceleration parameter $(q)$ and Hubble parameter ($H$) involved with bulk viscous coefficients $\xi_0$, $\xi_1$ and $\xi_2$ for the different ranges of $\lambda$ and $c_1$ in different types of the universe for $\gamma=0$, $\gamma=\frac{2}{3}$, $\gamma=1$ and $\gamma=\frac{4}{3}$.
\section{Discussion and conclusions}
The following observations are made from table-1 (for $\gamma=0 (p+\rho=0)$):
\begin{itemize}
  \item The deceleration parameter $q=-1$ throughout the evolution and the Hubble parameter $H$ is negative throughout the evolution for $c_1=0, \xi_0>0, \xi_2>0$ and, no restrictions for $\xi_1$ and $\lambda$, so that the universe is contracting and accelerating exponentially for $c_1=0, \xi_0>0, \xi_1+\xi_2<0$ and for all values of $\lambda$.
\item The deceleration parameter $q=-1$ and the Hubble parameter $H$ is positive throughout the evolution for $c_1=0, \xi_0>0, \xi_1+\xi_2<0$ and for all values of $\lambda$, so that the universe is expanding and accelerating exponentially for $c_1=0, \xi_0>0, \xi_1+\xi_2<0$ and for all values of $\lambda$.
\item For $c_1>0, \lambda>0, -4\pi<\lambda<\frac{1}{3\xi_2}-4\pi, \xi_0>0, \xi_2\ge \frac{1}{12\pi}$ and $0<\xi_2<\frac{1}{12\pi}$, the deceleration parameter $q<-1$ for small or present time and $q\rightarrow -1$ for large t, i. e. $t\rightarrow \infty$, and the Hubble parameter $H$ is varies from positive to negative as time increases, so the universe is expanding and accelerating in supper exponential way to contracting in exponential way.
\item The universe is expanding and accelerating in supper exponential way to expanding and accelerating in exponential way as the deceleration parameter $q<-1$ for present time and $q\rightarrow-1$ as $t\rightarrow\infty$, and the Hubble parameter is positive throughout the evolution for $c_1>0, \lambda>0, \xi_1+\xi_2<0$, $0<\xi_2<\frac{1}{12\pi}$, $-4\pi<\lambda<\frac{1}{3\xi_2}-4\pi$ and $\xi_2\ge \frac{1}{12\pi}$.
\item The universe is expanding and accelerating to expanding and decelerating as the deceleration parameter $q$ and the Hubble parameter $H$ varies from negative to positive for $c_1>0, \lambda>0, \xi_0>0, \xi_2>\frac{2}{3(8\pi+2\lambda)}$ and $\xi_1+\xi_2>0$.
\item The universe is expanding and and accelerating in supper exponentially to expanding and decelerating in standard way as the deceleration parameter $q$ varies from negative to positive and the Hubble parameter $H$ is positive throughout the evolution for $c_1>0, \lambda>0$ and $\xi_1+\xi_2<0$.
\item For $c_1>0, \lambda<-4\pi, \xi_0>0$ and $\xi_2>0$, the deceleration parameter varies from negative to positive and the Hubble parameter varies from positive to zero, so the universe is expanding and accelerating in super exponential way to accelerating in standard way.
\item The universe is accelerating and contracting to accelerating in standard way as the deceleration parameter is negative throughout the evolution and the Hubble parameter varies from negative to zero for $c_1<0, \lambda>0, \xi_0>0, \xi_2>\frac{2}{3(8\pi+2\lambda)}$, $-4\pi<\lambda<0$, $\lambda<-4\pi$ and $\xi_2>0$.
\item The universe is decelerating and contracting to accelerating and contracting in standard way as the deceleration parameter varies from positive to negative and $q\rightarrow -1$ as $t\rightarrow \infty$, and the Hubble parameter is negative throughout the evolution for $c_1<0, \lambda>0, \xi_0>0, 0<\xi_2<\frac{2}{3(8\pi+2\lambda)}$, $-4\pi<\lambda<0$.

\end{itemize}
The following observations are made from table-2 (for $\gamma=\frac{2}{3} (p+\frac{\rho}{3}=0)$):
\begin{itemize}
  \item The universe is accelerating and contracting to accelerating and expanding exponentially and vise versa as the deceleration parameter $q=-1$ and the Hubble parameter $H$ varies from negative to positive for some $\lambda$ and positive to negative for some $\lambda$, when $c_1=0, \xi_0>0, \xi_1>0, \xi_2>0$ and for all $\lambda$.
\item When $c_1>0, \lambda>-2.4\pi, \xi_0>0, \xi_1>0$ and $0<\xi_2<\frac{1}{2(\lambda+4\pi)(\lambda+2\pi)(24\pi+10\lambda)}$, the deceleration parameter $q$ is negative and $q\rightarrow -1$ as $t\rightarrow\infty$, and the Hubble parameter $H$ is positive throughout the evolution, so the universe is expanding and accelerating in supper exponentially to expanding and accelerating exponentially.
\item When $c_1>0, -4\pi<\lambda<-2.4\pi, \xi_0>0, \xi_1>0$ and $0<\xi_2<\frac{1}{2(\lambda+4\pi)(\lambda+2\pi)(24\pi+10\lambda)}$, the deceleration parameter varies from negative to positive and the Hubble parameter varies from negative to zero, so the universe is accelerating and contracting in standard way.
\item When $c_1>0, \lambda<-4\pi, \xi_0>0, \xi_1>0$ and $\xi_2>0$, the deceleration parameter $q<-1$ throughout the evolution and the Hubble parameter $H$ is positive throughout the evolution, so the universe is expanding and accelerating in supper exponential way.
\item When $c_1<0, \lambda>-2.4\pi, \xi_0>0, \xi_1>0$ and $0<\xi_2<\frac{1}{2(\lambda+4\pi)(\lambda+2\pi)(24\pi+10\lambda)}$, the deceleration parameter $q$ varies from positive to $-1$ as $t\rightarrow\infty$ and the Hubble parameter $H$ is positive throughout the evolution, so the universe is decelerating and expanding to accelerating and expanding in exponential way.
\item When $c_1<0, -4\pi<\lambda<-2.4\pi, \xi_0>0, \xi_1>0$ and $0<\xi_2<\frac{1}{2(\lambda+4\pi)(\lambda+2\pi)(24\pi+10\lambda)}$, the deceleration parameter is positive throughout the evolution and the Hubble parameter is varies from negative to zero, so the universe is decelerating and contracting in standard way.
\item When $c_1<0, \lambda<-4\pi, \xi_0>0, \xi_1>0$ and $\xi_2>0$, the deceleration parameter $q$ varies from positive to $q<-1$ for large $t$ and the Hubble parameter $H$ is positive throughout the evolution, so the universe is decelerating and expanding to accelerating in supper exponential way.
\end{itemize}
The following observations are made from table-3 (for $\gamma=1 (p=0)$):
\begin{itemize}
  \item The universe is accelerating and expanding in exponential way as the deceleration parameter $q=-1$ and the Hubble parameter is positive throughout the evolution for $c_1=0, \xi_0>0, \xi_1+\xi_2\rightarrow 0 ~~\mbox{or}~~\xi_1+\xi_2\le0 ~~ \mbox{and}~~0\le\xi_2<\frac{2(8\pi+3\lambda)}{3(8\pi+2\lambda)(8\pi+4\lambda)}$.
\item The universe is accelerating and expanding in supper exponential way to expanding and accelerating in exponential way as the deceleration parameter $q$ varies from $q<-1$ to $q=-1$ as $t\rightarrow\infty$ and the Hubble parameter is positive for $c_1>0, \lambda\ge0, \lambda\le -4\pi, \xi_0>0, \xi_1+\xi_2\rightarrow 0 ~~\mbox{or}~~\xi_1+\xi_2\le0 ~~ \mbox{and}~~0\le\xi_2<\frac{2(8\pi+3\lambda)}{3(8\pi+2\lambda)(8\pi+4\lambda)}$.
\item When $c_1>0, -4\pi<\lambda<0, \xi_0>0, \xi_1+\xi_2\rightarrow 0 ~~\mbox{or}~~\xi_1+\xi_2\le0 ~~ \mbox{and}~~0\le\xi_2<\frac{2(8\pi+3\lambda)}{3(8\pi+2\lambda)(8\pi+4\lambda)}$, the deceleration parameter $q<-1$ and the Hubble parameter is positive, so the universe is accelerating and expanding in supper exponential way.
\item $c_1<0, \lambda\ge0, \lambda\le -4\pi, \xi_0>0, \xi_1+\xi_2\rightarrow 0 ~~\mbox{or}~~\xi_1+\xi_2\le0 ~~ \mbox{and}~~0\le\xi_2<\frac{2(8\pi+3\lambda)}{3(8\pi+2\lambda)(8\pi+4\lambda)}$, the deceleration parameter varies from negative to $q=-1$ as $t\rightarrow\infty$ and the Hubble parameter is positive, so the the universe is accelerating and expanding in standard way to expanding and accelerating in exponential way.
\item When $c_1<0, -4\pi<\lambda<0, \xi_0>0, \xi_1+\xi_2\rightarrow 0 ~~\mbox{or}~~\xi_1+\xi_2\le0 ~~ \mbox{and}~~0\le\xi_2<\frac{2(8\pi+3\lambda)}{3(8\pi+2\lambda)(8\pi+4\lambda)}$, the deceleration parameter varies from negative to positive and the Hubble parameter is positive, so the universe is accelerating and expanding in standard way to decelerate in standard way.
\end{itemize}
The following observations are made from table-4 (for $\gamma=\frac{4}{3} (p=\frac{\rho}{3})$):
\begin{itemize}
  \item The universe is accelerating and expanding exponentially as the deceleration parameter $q=-1$ and the Hubble parameter $H$ is positive throughout the evolution for $c_1=0, \xi_0>0, \xi_1+\xi_2\rightarrow 0 ~~\mbox{or}~~\xi_1+\xi_2\le0 ~~ \mbox{and}~~ 0\le\xi_2<\frac{2(24\pi+8\lambda)}{3(8\pi+2\lambda)(8\pi+4\lambda)}$.
\item The universe is accelerating and expanding in supper exponential way to expanding and accelerating in exponential way as the deceleration parameter varies from $q<-1$ to $q=-1$ as $t\rightarrow\infty$ and the Hubble parameter is positive for $c_1>0, \lambda>-3\pi, \xi_0>0, \xi_1>0, \xi_1+\xi_2\rightarrow 0 ~~\mbox{or}~~\xi_1+\xi_2\le0 ~~ \mbox{and}~~ 0\le\xi_2<\frac{2(24\pi+8\lambda)}{3(8\pi+2\lambda)(8\pi+4\lambda)}$.
\item When $c_1>0, -4\pi<\lambda <-3\pi, \xi_0>0, \xi_1>0, \xi_1+\xi_2\rightarrow 0 ~~\mbox{or}~~\xi_1+\xi_2\le0, 0\le\xi_2<\frac{2(24\pi+8\lambda)}{3(8\pi+2\lambda)(8\pi+4\lambda)}$ and $\lambda<-4\pi, \xi_0>0, \xi_1>0, \xi_2>0$, the deceleration parameter varies from $q=-1$ to $q<-1$ as $t\rightarrow\infty$ and the Hubble parameter varies from positive to zero, so the universe is accelerating and expanding in exponential way to accelerating and expanding in supper exponential way.
\item When $c_1<0, \lambda>-3\pi, \xi_0>0, \xi_1>0, \xi_1+\xi_2\rightarrow 0 ~~\mbox{or}~~\xi_1+\xi_2\le0 ~~ \mbox{and}~~ 0\le\xi_2<\frac{2(24\pi+8\lambda)}{3(8\pi+2\lambda)(8\pi+4\lambda)}$, the deceleration parameter varies from $-1<q<0$ to $q=-1$ as $t\rightarrow\infty$ and the Hubble parameter is positive, so the universe is accelerating and expanding in standard way to expanding and accelerating in exponential way.
\item When $c_1<0, -4\pi<\lambda<-3\pi, \xi_0>0, \xi_1>0, \xi_1+\xi_2\rightarrow 0 ~~\mbox{or}~~\xi_1+\xi_2\le0 ~~ \mbox{and}~~ 0\le\xi_2<\frac{2(24\pi+8\lambda)}{3(8\pi+2\lambda)(8\pi+4\lambda)}$, the deceleration parameter varies from $-1<q<0$ to $q>0$ as $t\rightarrow\infty$ and the Hubble parameter varies from positive to zero, so the universe is accelerating and expanding in standard way to decelerating and expanding.
\item When $c_1<0, \lambda<-4\pi, \xi_0>0, \xi_1>0, \xi_2>0$, the deceleration parameter $q=-1$ to $q>0$ as $t\rightarrow\infty$ and the Hubble parameter is varies from positive to zero, so the universe is accelerating and expanding in exponential way to decelerating in standard way.

\end{itemize}
In this article, we carried out a study of the bulk viscous matter dominated universe with the bulk viscosity coefficient $\xi=\xi_0+\xi_1\frac{\dot{a}}{a}+\xi_2\frac{\ddot{a}}{\dot{a}}$ within the frame work of $f(R, T)$ gravity.
In this work, the model proposed by Avelino and Nucamendi \cite{Avelino} has been extended
and improved upon to reflect the more general situation. We extend their work into $f(R, T)$ gravity and the bulk viscous coefficient is proportional to the linear combination of three terms,
such as $\xi=\xi_0+\xi_1\frac{\dot{a}}{a}+\xi_2\frac{\ddot{a}}{\dot{a}}$, (where $\xi_0$, $\xi_1$ and $\xi_2$ are the positive constants) rather than
$\xi=\xi_0+\xi_1\frac{\dot{a}}{a}$.
The bulk viscous matter simultaneously represents dark matter and dark energy and causes the recent acceleration, therefore this model solves coincidence problem automatically. The bulk viscous fluid is a viable candidate to explain early and late time expansion of the universe. Therefore, in this article we explored the evolution of the universe driven by a kind of viscous fluid by assuming general form of bulk viscous coefficients. We discussed the expansion history of the universe with viscosity. We discussed the various phases and their possible transitions for all possible range of $\lambda, c_1$ with $\xi_0, \xi_1$ and $\xi_2$. We obtained time dependent deceleration parameter $q$ and Hubble parameter $H$ which describe the decelerated/accelerated and transition from decelerated/accelerated to accelerated/decelerated phase. The existence of sudden singularities to the Friedmann universes of higher-order lagrangian theories of gravity discussed by \cite{Barrow65}. The big-rip singularity is called the type I singularity \cite{Tsujikawa}, the behavior of type I singularity is as follows: $a\rightarrow\infty, ~\rho\rightarrow\infty,~|p|\rightarrow\infty,$ as $t\rightarrow t_s$. Finally, we observed big-rip or type I singularity at
$t=\frac{\bigg[2-3(8\pi+2\lambda)\left(\frac{\gamma\lambda}{(8\pi+4\lambda-\lambda\gamma)}+1\right)\xi_2\bigg]}
{\bigg[\frac{3\lambda\xi_0\gamma(8\pi+2\lambda)}{(8\pi+4\lambda-\lambda\gamma)}+3\xi_0(8\pi+2\lambda)\bigg]}\times \ln\bigg[c_1
\frac{\bigg[\frac{3\lambda\xi_0\gamma(8\pi+2\lambda)}{(8\pi+4\lambda-\lambda\gamma)}+3\xi_0(8\pi+2\lambda)\bigg]}
{\bigg[\frac{(8\pi+2\lambda)3\gamma}{8\pi+4\lambda-\lambda\gamma}(1-\lambda(\xi_1+\xi_2))-3(8\pi+2\lambda)(\xi_1+\xi_2)\bigg]}\bigg]$ in our model.
In conclusion, the authors strongly emphasize that perfect fluid is just a limiting case of a general bulk viscous medium that is more practical in the astrophysical sense. Therefore, it is meaningful to study the early and late time cosmic evolution of the universe with the bulk viscous fluid in $f(R, T)$ gravity, which describes the evolution of the universe in various ways for different range of $\lambda$, $\xi_0$, $\xi_1$ and $\xi_2$.
Subsequently, our paper provides an interesting topic for the further study of such kind of important cosmological models, where the matter content of the universe is filled with bulk viscous fluid for variable bulk viscous coefficients, and at the same time, the following problems
can be considered in our future research work.
\begin{enumerate}
  \item  The same work may extend to Kaluza-Klein theory.
\item  In this paper, authors have taken $f(R, T)=R+2f(T)$, where $f(T)=\lambda T$. The researcher may think the different form of $f(R, T)$ and $f(T)$, say instead of linear, may think about quadratic form.
\item We may look towards the validity of the second law of thermodynamics in the presence of the bulk viscous fluid with variable coefficients.
\item In this paper, authors have taken the matter part of the universe filled with imperfect fluid within the frame work of isotropic space time, the researcher may extend this work within the framework of anisotropic space time (viz. Bianchi type space time).
\item The double special relativity generalized to curved space-time, and this doubly general theory of relativity is called gravity’s rainbow \cite{Magueijo}. In this theory, the geometry of space-time depends on the energy of the test particle.
Therefore, the geometry of space-time is represented by a family of energy dependent metrics forming a rainbow of metrics.
This is the reason the theory is called gravity’s rainbow. The gravity’s rainbow is extensively studied in order
to explore various aspects for black holes and cosmology (\cite{Hendi}, \cite{Chatrabhuti}, \cite{Gangopadhyay}, \cite{Hendi2}, \cite{Ali}, \cite{Ali1}, \cite{Hendi3}, \cite{Rudra}, \cite{Hendi4}, \cite{Ashour}, \cite{Hendi5}, \cite{Gima}). Therefore, it is more interesting to study our work in gravity's rainbow.
\end{enumerate}
\textbf{Acknowledgments}: The authors thank the anonymous referees
for enlightening comments and suggestions which
substantially improved the quality of the research work.

\begin{table}
  \centering
  \caption{For $\gamma=0 (p+\rho=0)$}\label{1}
  \begin{tabular}{|cccccc|}
  \hline\noalign{\smallskip}
\makecell{ Range \\of $c_1$} & \makecell{Range \\of $\lambda$} & \makecell{ Constraints of \\bulk coefficients \\$\xi_0, \xi_1$ and  $\xi_2$ } & \makecell{Variation\\ of $q$} & \makecell{Variation \\of $H$} & \makecell{Evolution \\of the \\universe}\\
\noalign{\smallskip}\hline\noalign{\smallskip}
    $c_1=0$ & for all $\lambda$ & \makecell{$\xi_0>0$, \\ $\xi_2>0$} & $q=-1$ & \makecell{negative \\throughout \\the evolution}& \makecell{contracting and\\ accelerating \\exponentially} \\ \hline
 $c_1=0$ & for all $\lambda$ & \makecell{$\xi_0>0$\\$\xi_1+\xi_2<0$} & $q=-1$ & \makecell{positive \\throughout \\the evolution}& \makecell{expanding and\\ accelerating \\exponentially} \\ \hline
 $c_1>0$ & $\lambda >0$ & \makecell{$\xi_0>0$,\\$0<\xi_2<\frac{1}{12\pi}$} & \makecell{$q<-1$ and \\$q\rightarrow -1$ \\as $t\rightarrow \infty$} & \makecell{positive to \\negative} & \makecell{expanding and \\accelerating  in \\supper exponentially \\to expanding and \\ contracting in \\exponentially} \\ \hline
 $c_1>0$ & $\lambda>0$ & \makecell{$\xi_1+\xi_2<0$,\\ $0<\xi_2<\frac{1}{12\pi}$} & \makecell{$q<-1$ and \\$q\rightarrow -1$ \\as $t\rightarrow \infty$} & \makecell{positive\\ throughout \\the evolution} & \makecell{expanding and \\accelerating  in \\supper exponentially \\to expanding and \\ accelerating in \\exponentially} \\ \hline
 $c_1>0$ & $\lambda>0$ & \makecell{$\xi_0>0$,\\ $\xi_2>\frac{2}{3(8\pi+2\lambda)}$,\\$\xi_1+\xi_2>0$} & \makecell{negative \\to positive} & \makecell{negative \\to positive} & \makecell{expanding and \\accelerating to \\expanding and \\ decelerating} \\ \hline
 $c_1>0$ & $\lambda>0$ & \makecell{$\xi_1+\xi_2<0$} & \makecell{negative \\to positive} & \makecell{positive \\throughout \\ the evolution} & \makecell{expanding and \\accelerating in \\super exponentially \\to expanding and \\ decelerating} \\ \hline
 $c_1>0$ & \makecell{$-4\pi<\lambda<\frac{1}{3\xi_2}-4\pi$} & \makecell{$\xi_0>0$, \\ $\xi_2\ge\frac{1}{12\pi}$ } & \makecell{$q<-1$ and \\$q\rightarrow -1$ \\as $t\rightarrow\infty$} & \makecell{positive to\\ negative} & \makecell{expanding and \\accelerating in \\super exponentially \\to accelerating \\and contracting}\\\hline
 $c_1>0$ & \makecell{$-4\pi<\lambda<\frac{1}{3\xi_2}-4\pi$} & \makecell{$\xi_1+\xi_2<0$, \\ $\xi_2\ge\frac{1}{12\pi}$ } & \makecell{$q<-1$ and \\$q\rightarrow -1$ \\as $t\rightarrow\infty$} & \makecell{positive \\throughout \\the evolution} & \makecell{expanding and \\accelerating in \\super exponentially \\to accelerating and \\ expanding in \\exponentially}\\\hline
 $c_1>0$ & $\lambda<-4\pi$ & \makecell{$\xi_0>0$,\\ $\xi_2>0$} & \makecell{negative \\to positive} & \makecell{positive \\to zero} & \makecell{expanding and \\accelerating in \\super exponentially \\to accelerating in \\ standard way}\\\hline
 %$c_1<0$ & $\lambda>0$ & \makecell{$\xi_0>0$, \\ $\xi_2>\frac{2}{3(8\pi+2\lambda)}$} & $q<0$ & \makecell{negative \\to zero} & \makecell{accelerating and\\contracting to \\accelerating in\\standard way} \\\hline
 \end{tabular}
\end{table}
 \newpage
 \begin{table}
  %\centering
  %\caption{For $\gamma=0$}\label{1}
  \begin{tabular}{|cccccc|}
  \hline\noalign{\smallskip}
  $c_1<0$ & $\lambda>0$ & \makecell{$\xi_0>0$, \\ $\xi_2>\frac{2}{3(8\pi+2\lambda)}$} & $q<0$ & \makecell{negative \\to zero} & \makecell{accelerating and\\contracting to \\accelerating in\\standard way} \\\hline
 $c_1<0$ & $\lambda>0$ & \makecell{$\xi_0>0$, \\ $0<\xi_2<\frac{2}{3(8\pi+2\lambda)}$} & \makecell{positive \\to negative \\and $q\rightarrow-1$\\as $t\rightarrow\infty$} & \makecell{negative \\throughout \\the evolution} & \makecell{decelerating and\\contracting to \\accelerating and\\contracting} \\\hline
 $c_1<0$ & $-4\pi<\lambda<0$ & \makecell{$\xi_0>0$, \\ $\xi_2>\frac{2}{3(8\pi+2\lambda)}$} & \makecell{$q$ is negative\\ throughout the\\evolution} & \makecell{negative \\to zero} & \makecell{accelerating and\\contracting to \\accelerating in\\standard way} \\\hline
 $c_1<0$ & $-4\pi<\lambda<0$ & \makecell{$\xi_0>0$, \\ $0<\xi_2<\frac{2}{3(8\pi+2\lambda)}$} & \makecell{$q$ is positive\\ to negative and\\
 $q\rightarrow -1 $ \\as $t\rightarrow\infty $} & \makecell{negative \\throughout \\the evolution} & \makecell{decelerating and\\contracting to \\accelerating and\\contracting} \\\hline
 $c_1<0$ & $\lambda<-4\pi$ & \makecell{$\xi_0>0$, \\ $\xi_2>0$} & \makecell{$q$ is negative\\throughout the\\evolution} & \makecell{negative \\to zero} & \makecell{accelerating and\\contracting to \\accelerating in\\standard way} \\\hline
  \end{tabular}
\end{table}

\begin{table}
  \centering
  \caption{For $\gamma=\frac{2}{3} (p+\frac{\rho}{3}=0)$}\label{2}
  \begin{tabular}{|cccccc|}
  \hline\noalign{\smallskip}
\makecell{ Range \\of $c_1$} & \makecell{Range \\of $\lambda$} & \makecell{ Constraints of \\bulk coefficients \\$\xi_0, \xi_1$ and  $\xi_2$ } & \makecell{Variation\\ of $q$} & \makecell{Variation \\of $H$} & \makecell{Evolution \\of the \\universe}\\
\noalign{\smallskip}\hline\noalign{\smallskip}
$c_1=0$ & \makecell{for all $\lambda$} & \makecell{$\xi_0>0, \xi_1>0,$\\$\xi_2>0$} & \makecell{$q=-1$} & \makecell{negative to \\positive for\\some $\lambda$ and\\positive to \\negative for\\some $\lambda$} & \makecell{accelerating and\\ contracting to \\accelerating and\\ expanding \\exponentially\\and vise versa}\\\hline
$c_1>0$ & \makecell{$\lambda>-2.4\pi$} & \makecell{$\xi_0>0, \xi_1>0,$\\$\xi_2<\frac{1}{2(\lambda+4\pi)(\lambda+2\pi)(24\pi+10\lambda)}$, \\$\xi_2>0$} & \makecell{negative\\and\\$q\rightarrow -1$ \\as $t\rightarrow \infty$} & \makecell{positive \\ throughout\\ the evolution} & \makecell{expanding and \\accelerating \\in super\\ exponentially to\\ expanding and \\accelerating\\exponentially}\\\hline

$c_1>0$ & $-4\pi<\lambda<-2.4\pi$ & \makecell{$\xi_0>0, \xi_1>0,$\\$\xi_2<\frac{1}{2(\lambda+4\pi)(\lambda+2\pi)(24\pi+10\lambda)}$ \\$\xi_2>0$} & \makecell{negative \\to positive} & \makecell{negative\\ to zero} & \makecell{accelerating\\ and contracting} \\\hline
$c_1>0$ & $\lambda<-4\pi$ & \makecell{$\xi_0>0, \xi_1>0$,\\$\xi_2>0$} & \makecell{$q<-1$ \\throughout\\ the evolution} & \makecell{positive\\throughout\\the evolution} & \makecell{expanding and\\ accelerating\\ in supper\\ exponentially}\\\hline

$c_1<0$ & \makecell{$\lambda>-2.4\pi$} & \makecell{$\xi_0>0, \xi_1>0,$\\$\xi_2<\frac{1}{2(\lambda+4\pi)(\lambda+2\pi)(24\pi+10\lambda)}$,\\$\xi_2>0$} & \makecell{positive\\and\\$q\rightarrow -1$ \\as $t\rightarrow \infty$} & \makecell{positive \\ throughout\\ the evolution} & \makecell{decelerating\\and expanding to\\accelerating\\and expanding\\exponentially}\\\hline

$c_1<0$ & $-4\pi<\lambda<-2.4\pi$ & \makecell{$\xi_0>0, \xi_1>0,$\\$\xi_2<\frac{1}{2(\lambda+4\pi)(\lambda+2\pi)(24\pi+10\lambda)}$,\\$\xi_2>0$} & \makecell{positive\\throughout\\the evolution} & \makecell{negative\\ to zero} & \makecell{decelerating\\ and contracting} \\\hline
$c_1<0$ & $\lambda<-4\pi$ & \makecell{$\xi_0>0, \xi_1>0$,\\$\xi_2>0$} & \makecell{positive\\to $q<-1$\\ for large\\$t$} & \makecell{positive\\throughout\\the evolution} & \makecell{decelerating\\and expanding\\to accelerating\\in supper\\exponentially}\\\hline
\end{tabular}
\end{table}

\begin{table}[ht!]
  %\centering
  \caption{For $\gamma=1 (p=0)$}\label{3}
  \begin{tabular}{|cccccc|}
  \hline\noalign{\smallskip}
\makecell{ Range \\of $c_1$} & \makecell{Range \\of $\lambda$} & \makecell{ Constraints of \\bulk coefficients \\$\xi_0, \xi_1$ and  $\xi_2$ } & \makecell{Variation\\ of $q$} & \makecell{Variation \\of $H$} & \makecell{Evolution \\of the \\universe}\\
\noalign{\smallskip}\hline\noalign{\smallskip}

$c_1=0$ & \makecell{for all $\lambda$} & \makecell{$\xi_0>0,$\\$0\le \xi_2<\frac{2(8\pi+3\lambda)}{3(8\pi+2\lambda)(8\pi+4\lambda)}$\\$\xi_1+\xi_2\rightarrow 0$\\or $\xi_1+\xi_2\le0$} & \makecell{$q=-1$\\throughout\\the evolution} & \makecell{positive} & \makecell{accelerating and\\ expanding \\exponentially}\\\hline
$c_1>0$ & \makecell{$\lambda\le -4\pi$, $\lambda\ge 0$} & \makecell{$\xi_0>0,$\\$0\le \xi_2<\frac{2(8\pi+3\lambda)}{3(8\pi+2\lambda)(8\pi+4\lambda)}$\\$\xi_1+\xi_2\rightarrow 0$\\or $\xi_1+\xi_2\le0$} & \makecell{$q<-1$ to\\$q=-1$ as\\
$t\rightarrow\infty$} & \makecell{positive} & \makecell{accelerating and\\expanding in\\supper exponential\\ way to expanding\\and accelerating\\in exponential way}\\\hline
$c_1>0$ & \makecell{$-4\pi<\lambda <0$} & \makecell{$\xi_0>0,$\\$0\le \xi_2<\frac{2(8\pi+3\lambda)}{3(8\pi+2\lambda)(8\pi+4\lambda)}$\\$\xi_1+\xi_2\rightarrow 0$\\or $\xi_1+\xi_2\le0$} & \makecell{$q<-1$} & \makecell{positive} & \makecell{accelerating and\\expanding in\\supper exponential\\ way}\\\hline
$c_1<0$ & \makecell{$\lambda\le -4\pi$, $\lambda\ge 0$} & \makecell{$\xi_0>0,$\\$0\le \xi_2<\frac{2(8\pi+3\lambda)}{3(8\pi+2\lambda)(8\pi+4\lambda)}$\\$\xi_1+\xi_2\rightarrow 0$\\or $\xi_1+\xi_2\le0$} & \makecell{negative to\\$q=-1$ as\\
$t\rightarrow\infty$} & \makecell{positive} & \makecell{accelerating and\\expanding in\\standard way\\ to expanding\\and accelerating\\in exponential way}\\\hline
$c_1<0$ & \makecell{$-4\pi<\lambda <0$} & \makecell{$\xi_0>0,$\\$0\le \xi_2<\frac{2(8\pi+3\lambda)}{3(8\pi+2\lambda)(8\pi+4\lambda)}$\\$\xi_1+\xi_2\rightarrow 0$\\or $\xi_1+\xi_2\le0$} & \makecell{negative\\to positive} & \makecell{positive} & \makecell{accelerating and\\expanding in\\standard way\\to decelerate\\in standard way}\\\hline

\end{tabular}
\end{table}

\begin{table}[ht!]
  %\centering
  \caption{For $\gamma=\frac{4}{3} (p=\frac{\rho}{3})$}\label{4}
  \begin{tabular}{|cccccc|}
  \hline\noalign{\smallskip}
\makecell{ Range \\of $c_1$} & \makecell{Range \\of $\lambda$} & \makecell{ Constraints of \\bulk coefficients \\$\xi_0, \xi_1$ and  $\xi_2$ } & \makecell{Variation\\ of $q$} & \makecell{Variation \\of $H$} & \makecell{Evolution \\of the \\universe}\\
\noalign{\smallskip}\hline\noalign{\smallskip}

$c_1=0$ & \makecell{for all $\lambda$} & \makecell{$\xi_0>0,$\\$0\le \xi_2<\frac{2(24\pi+8\lambda)}{3(8\pi+2\lambda)(8\pi+4\lambda)}$\\$\xi_1+\xi_2\rightarrow 0$\\or $\xi_1+\xi_2\le0$} & \makecell{$q=-1$\\throughout\\the evolution} & \makecell{positive} & \makecell{accelerating and\\ expanding \\exponentially}\\\hline
$c_1>0$ & \makecell{$\lambda> -3\pi$} & \makecell{$\xi_0>0, \xi_1>0$\\$0\le \xi_2<\frac{2(24\pi+8\lambda)}{3(8\pi+2\lambda)(8\pi+4\lambda)}$\\$\xi_1+\xi_2\rightarrow 0$\\or $\xi_1+\xi_2\le0$} & \makecell{$q<-1$ to\\$q=-1$ as\\
$t\rightarrow\infty$} & \makecell{positive} & \makecell{accelerating and\\expanding in\\supper exponential\\ way to expanding\\and accelerating\\in exponential way}\\\hline
$c_1>0$ & \makecell{$-4\pi<\lambda <-3\pi$} & \makecell{$\xi_0>0, \xi_1>0$\\$0\le \xi_2<\frac{2(24\pi+8\lambda)}{3(8\pi+2\lambda)(8\pi+4\lambda)}$\\$\xi_1+\xi_2\rightarrow 0$\\or $\xi_1+\xi_2\le0$} & \makecell{$q=-1$ to\\$q<-1$} & \makecell{positive\\ to zero} & \makecell{accelerating and\\expanding in\\ exponentially to\\ accelerating and\\ expanding in supper\\ exponential way}\\\hline
$c_1>0$ & \makecell{$\lambda<-4\pi$} & \makecell{$\xi_0>0, \xi_1>0$\\$\xi_2>0$} & \makecell{$q=-1$ to\\$q<-1$ as\\$t\rightarrow\infty$} & \makecell{positive\\ to zero} & \makecell{accelerating\\ and expanding \\exponentially to \\accelerating\\ and expanding \\in supper \\exponentially}\\\hline
$c_1<0$ & \makecell{$\lambda> -3\pi$} & \makecell{$\xi_0>0, \xi_1>0$\\$0\le \xi_2<\frac{2(24\pi+8\lambda)}{3(8\pi+2\lambda)(8\pi+4\lambda)}$\\$\xi_1+\xi_2\rightarrow 0$\\or $\xi_1+\xi_2\le0$} & \makecell{$-1<q<0$ to\\$q=-1$ as\\
$t\rightarrow\infty$} & \makecell{positive} & \makecell{accelerating and\\expanding in\\standard way\\ to expanding\\and accelerating\\in exponential way}\\\hline
$c_1<0$ & \makecell{$-4\pi<\lambda <-3\pi$} & \makecell{$\xi_0>0, \xi_1>0$\\$0\le \xi_2<\frac{2(24\pi+8\lambda)}{3(8\pi+2\lambda)(8\pi+4\lambda)}$\\$\xi_1+\xi_2\rightarrow 0$\\or $\xi_1+\xi_2\le0$} & \makecell{$-1<q<0$ to\\$q>0$} & \makecell{positive\\ to zero} & \makecell{accelerating and\\expanding in\\ standard way to\\ decelerating and\\ expanding}\\\hline
$c_1<0$ & \makecell{$\lambda<-4\pi$} & \makecell{$\xi_0>0, \xi_1>0$\\$\xi_2>0$} & \makecell{$q=-1$ to\\$q>0$ as\\$t\rightarrow\infty$} & \makecell{positive\\ to zero} & \makecell{accelerating\\ and expanding \\exponentially to \\decelerating\\ in standard way}\\\hline
\end{tabular}
\end{table}

\end{document}